# A simple way of unifying the formulas for the Coulomb's law and Newton's law of the universal gravitation: An approach based on membranes

**Lizandro B. R. Zegarra[1, a], Milton C. Gutierrez[2], Fidel. A. V. Obeso[1], Luis T. Quispe[3] and L. E. G. Armas[4, b]**

[1]Departamento de Matemática de la Universidad Nacional del Santa – UNS, 508, Chimbote – Perú

[2]Departamento de Matemática de la Universidad Nacional de Trujillo – UNT, Trujillo – Perú

[3]Laboratorio de Películas Delgadas, Escuela Profesional de Física – Universidad Nacional de San Agustín de Arequipa-UNSA, Arequipa, Perú

[4]Grupo de Óptica Micro e Nanofabricação de Dispositivos (GOMNDI), Universidade Federal do Pampa, Campus Alegrete, 97546-550, Rio Grande do Sul - RS, Brazil

*Corresponding Authors*: Lizandro B. R. Zegarra
[a] e-mail: lreynazegarra@gmail.com

Luis E. G. Armas
[b] e-mail: legarmas@gmail.com , *Phonenumber* +55-55-34218400

**Abstract**

In this work, a new approach is presented with the aim of showing a simple way of unifying the classical formulas for the forces of the Coulomb's law of electrostatic interaction $(F_C)$ and the Newton's law of universal gravitation $(F_G)$. In this approach, these two forces are of the same nature and are ascribed to the interaction between two membranes that oscillate according to different curvature functions with the same spatial period $\xi\pi/k$ where $\xi$ is a dimensionless parameter and $k$ a wave number. Both curvature functions are solutions of the classical wave equation with wavelength given by the de Broglie relation. This new formula still keeps itself as the inverse square law, and it is like $F_C$ when the dimensionless parameter $\xi = 274$ and like $F_G$ when $\xi = 1.14198 \times 10^{45}$. It was found that the values of the parameter $\xi$ quantize the formula from which $F_C$ and $F_G$ are obtained as particular cases.

**Résumé**- Dans ce travail, une nouvelle approche est présentée dans le but de montrer une manière simple d'unifier les formules classiques des forces de la loi de Coulomb d'interaction électrostatique $(F_C)$ et de la loi de Newton de la gravitation universelle $(F_G)$. Dans cette approche, ces deux forces sont de même nature et sont attribuées à l'interaction entre deux membranes qui oscillent selon différentes fonctions de courbure avec la même période spatiale $\xi\pi/k$ où $\xi$ est un paramètre sans dimension et $k$ un nombre d'onde. Les deux fonctions de courbure sont des solutions de l'équation d'onde classique avec une longueur d'onde donnée par la relation de Broglie. Cette nouvelle formule se conserve toujours comme la loi des carrés inverse, et elle est comme $F_C$ lorsque le paramètre sans dimensión $\xi = 274$ et comme $F_G$ lorsque $\xi = 1.14198 \times 10^{45}$. On a constaté que les valeurs du paramètre $\xi$ quantifient la formule à partir de laquelle $F_C$ et $F_G$ sont obtenus comme cas particuliers.

## 1. Introduction

In nature, four types of fundamental interaction are known as described by Nagashima.[1] The fact that there are four independent and apparently unrelated fields of interaction is somewhat unsatisfactory and physicists like Einstein have speculated that the different interactions are different aspects of a single unified field. Gravity and Electromagnetism are the only two long range fundamental forces in the universe we know of, and their presence is ubiquitous.

Since Einstein's time, many scientist have attempted the unification of the classical theory of these two forces, attempts that until now continue several authors, such as: Corben[2] asserts that while the general theory of relativity has had phenomenal success in describing certain well-known effects associated with gravitational interaction and even with the path of a ray of light in a gravitational field, it has not been possible to obtain a thoroughly satisfactory unified theory of gravitational and electromagnetic phenomena. According to Fradkin and Tseytlin,[3] a central problem of modern high-energy physics is the unification of gravity with all the other fundamental interactions which would be consistent at the quantum level. Davidson and Owen,[4] have unified the Newton and the Coulomb laws via a generalized Reissner-Nordstrom geometry, with the electric charge resembling a topologically- frozen boost. Brandenburg[5] asserts that the gravity-electromagnetism theory uses concepts familiar to plasma physicists to produce a unification of gravity and electromagnetism. Assis,[6] by applying the generalized Weber's law for electromagnetism with the fourth-order terms and greater by $1/c$ to the force between two neutral dipoles, obtains an equivalent to Newton's law of universal gravitation as a fourth order electromagnetic effect, that is, he showed the possibility of deriving the gravitation from electromagnetism. Ghose[7,8] has shown that unification of gravity and electromagnetism can be achieved using an affine non-symmetric connection. This unification is based on projective invariance which is broken by matter fields, opening up the possibility of a unified theory of all forces in which gravity emerges as a classical field. Haug[9] has shown that the formula corresponding to the Newton's gravitational interaction and the formula corresponding to the Coulomb's electrostatic interaction are exactly the same formula on the Planck scale, and that naturally these two formulas calculate the same force only when considering two masses of Planck. For other masses that are not from Planck it is not true. Hendi and Sheykhi[10] argued that if both the gravitational and electromagnetic interaction were obtained from a holographic principle, they could be considered as a form of unification of the gravitational and electromagnetic force. They deduced that at very small distances, these two fundamental forces have the same behavior.

In this work, we show a simple way to unify the Coulomb's electrostatic interaction force $(F_C)$ and Newton's gravitational interaction force $(F_G)$, using an approach based on membranes, which are spatially two-dimensional surfaces characterized by their curvatures. It is worth to emphasize that the objective of this work is only to show that both forces can be written through a single formula, which depends of a dimensionless parameter $(\xi)$ and not of the classical constants $K_e$ (Coulomb's law) and $G$ (Newton's law), using an approach based on membranes. This approach is an idea inspired in string theory that consists in replacing the strings of string theory by certain membranes which don´t belong to any "membrane theory" we know, but that they behave under certain mechanism we describe below. For this purpose a basic definition of surfaces and gaussian curvature of a sphere and torus, as well as, wave equation is first introduced; following the mathematical development for the simple unification of Coulomb's and Newton's Forces.

**2. Theory**

**2.1** *Surfaces and gaussian curvature*

In this work, the notion of surface (membrane) is approached considering arguments of differential geometry. Roughly speaking, a surface in $R^3$ is obtained by taking pieces of a plane, deforming them and arranging them in such a way that the resulting figure has no sharp points, edges, or self-intersections and so that it makes sense to speak of a tangent plane at any points of the figure. This surface is smooth enough so that the usual notions of calculus can be extended to it. On the other hand, the gaussian curvature of a surface tell us how rapidly a surface pulls away from the tangent plane.[11]

We consider the surface of a sphere, the surface of a torus and the concept of curvature fundamentally, because the curvature of the sphere is like the inverse square law and the curvature of the torus leads to the inverse square law too. If we want to obtain some type of interaction, we need two entities at least. In this case we consider only two.

The gaussian curvature of a sphere, shown on Fig. 1(a), of radius $r$ is given by

$$K_S = \frac{1}{r^2} \tag{1}$$

so this is always positive and constant for a given value of $r$. On the other hand, the gaussian curvature of the torus $(K_T)$ shown in Fig. 1(b), is given by

$$K_T = \frac{\cos u}{R(a + R\cos u)} \qquad (2)$$

From this expression, it follows that $K_T = 0$ along the parallels $u = \frac{\pi}{2}$ and $u = \frac{3\pi}{2}$, $K_T < 0$ in the region of the torus given by $\frac{\pi}{2} < u < \frac{3\pi}{2}$ (notice that $a > R > 0$) and $K_T > 0$ in the region given by $0 < u < \frac{\pi}{2}$ or $\frac{3\pi}{2} < u < 2\pi$.

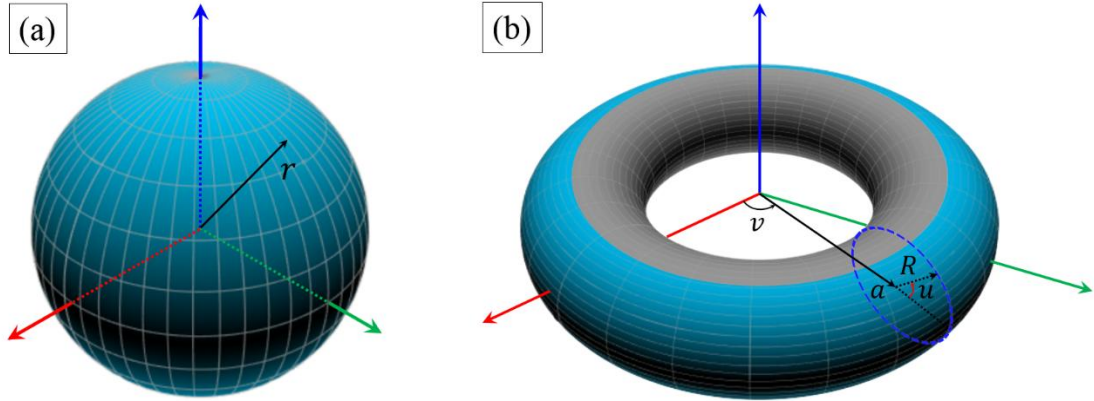

**FIG. 1.** Graphical representation of the surfaces (membranes) considered in this work. (a) Spherical surface of radius $r$, in which the gaussian curvature is always positive and (b) toroidal surface, in which the gaussian curvature can be negative (gray region), positive (light blue region) and zero (gray and light blue boundary).

In the development of the theory that will allow us to unify Newton and Coulomb laws, these two kinds of surfaces will be called membranes which are fixed in each point along a spatial line and whose respective curvatures as function of time are oscillating. Instead of a closed string and an open string we have a closed membrane (sphere) and an open membrane (part of the toroidal surface with negative gaussian curvature), respectively. These membranes can vibrate and interact between them, as will be described later.

### 2.2. The wave equation and traveling waves

The reason for the partial success of string theory is that it utilizes wave theory and resonant frequencies. Since in the development of this work, the idea of membranes is inspired on the strings, the classical wave equation is considered. Of interest are those waves that move freely through the medium, carry energy to other places in space and which are solutions of the classical wave equation. The interest of this kind of waves is because they are similar to those waves that describe the electrical or magnetic part of the electromagnetic waves. Furthermore, according to the quantum field theory the interaction forces are due to exchange of virtual particles. In the electromagnetic interaction case, this interaction is due to exchange of photons. However, due to wave – particle duality, and with the aim of unifying the Newton and Coulomb laws, we consider anyway instead of photons, waves. The curvatures of the membranes given by Eqs. (1) and (2) are given from the point of view of differential geometry and are statics. In order that these curvatures become dynamics, we further suppose that these ones are oscillating according to the solutions of the classical wave equation,

$$\frac{\partial^2 K}{\partial t^2} = v^2 \frac{\partial^2 K}{\partial x^2} \qquad (3a)$$

Solutions that we consider are given by

$$K = K_0 \sin^2(kx - \omega t) \qquad (3b)$$

where, $K$ is the gaussian curvature function; $K_0$ the amplitude of $K$; $k$ the angular spatial frequency of the wave; $\omega$ the angular temporal frequency and $x, t$ are the spatial and temporal coordinates. $K$ is considered of this form because the curvatures of the membranes, here considered, are always positive in one membrane and negative on the other one, so we ascribe the sign of them to the amplitude $K_0$, and for their oscillations the function $\sin^2(kx - \omega t)$, which is always positive. In this way, the form of Eq. (3b) describes the oscillation of the curvatures $K$ of the membranes, as we wish for the development of this work and it will be shown in the next section.

### 3. A new hypothesis: the behavior of membranes

We adopt that the membranes considered above satisfy the wave equation and oscillate with the wavelength given by the de Broglie relation. These membranes while evolve with time originate a set of concentric membranes. By considering the membranes given in Fig. 1, when $r = R$, a set of spheres are put together around the torus where its curvature is negative, as indicated by the arrows in Fig. 2(a). Then, we give the following definitions.

**Definition 1:** It is said that a topological change between a sets of points that form different surfaces happens, when because of the union of these sets of points, a new set of points that determines a new surface is obtained with a well-defined curvature. From Fig. 2(b) the simplest topological change obtained is that which gives as a result the surface of a torus. When the sphere oscillates according to the change of its radius with time, causes that the two − dimensional spatial surface traces out a three − dimensional volume, that is, two − dimensional spatial and one − dimensional temporal. But, due to Eq. (1), the change in the radius with time is the same to change the gaussian curvature $K$ with time.

Now, suppose that a topological change happened. Furthermore, suppose that the torus, as a result of the topological change, deforms to a sphere continuously while $a \to 0$. Then, the following definition turns out to be convenient.

**Definition 2:** An annihilation process of a torus to a sphere is that for which $\frac{a}{R} \to 0$ when a topological change has happened, as shown in Fig. 2(c).

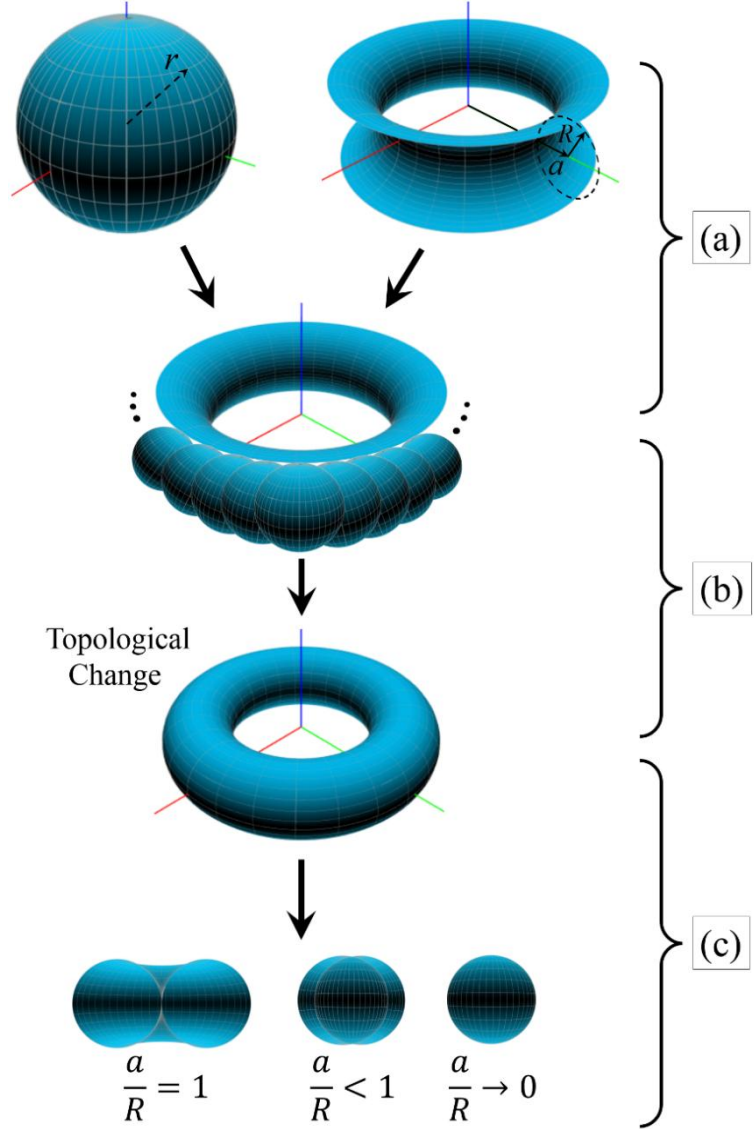

**FIG. 2.** (a) Union of a set of spheres around the torus where its curvature is negative. (b) Topological change to a torus minus the line (circumference) of zero curvature. (c) Annihilation process of the torus to a sphere. The figure on the left is such that each pair of circumferences obtained by cutting the torus with a plane $\left(\frac{a}{R}=1\right)$, will intersect $\left(\frac{a}{R}<1\right)$ and eventually overlap as $\frac{a}{R}\to 0$.

Figure 3 shows the behavior of the membranes after the annihilation process has occurred. That is, Fig. 3(a) shows the graphical representation of the curvature function $K = K_0(0)\sin^2(k(x,0)x)$ when $t=0$. The squared sine function gives the curvature of each fixed sphere in each fixed point of the spatial line $x$. Similarly, Fig. 3(b) shows the graphical representation of the curvature function $K = K_0(0)\sin^2(k(x,t)x - \omega(x,t)t)$ when $t \neq 0$. The squared sine function gives the curvatures of each of the set of concentric spheres that arise when $t$ varies, but that they are fixed at each point of the spatial line $x$. Both the curvature, frequency and wavenumber of each one of the concentric spheres (Fig. 3b) at each point $x$ change with time $t$.

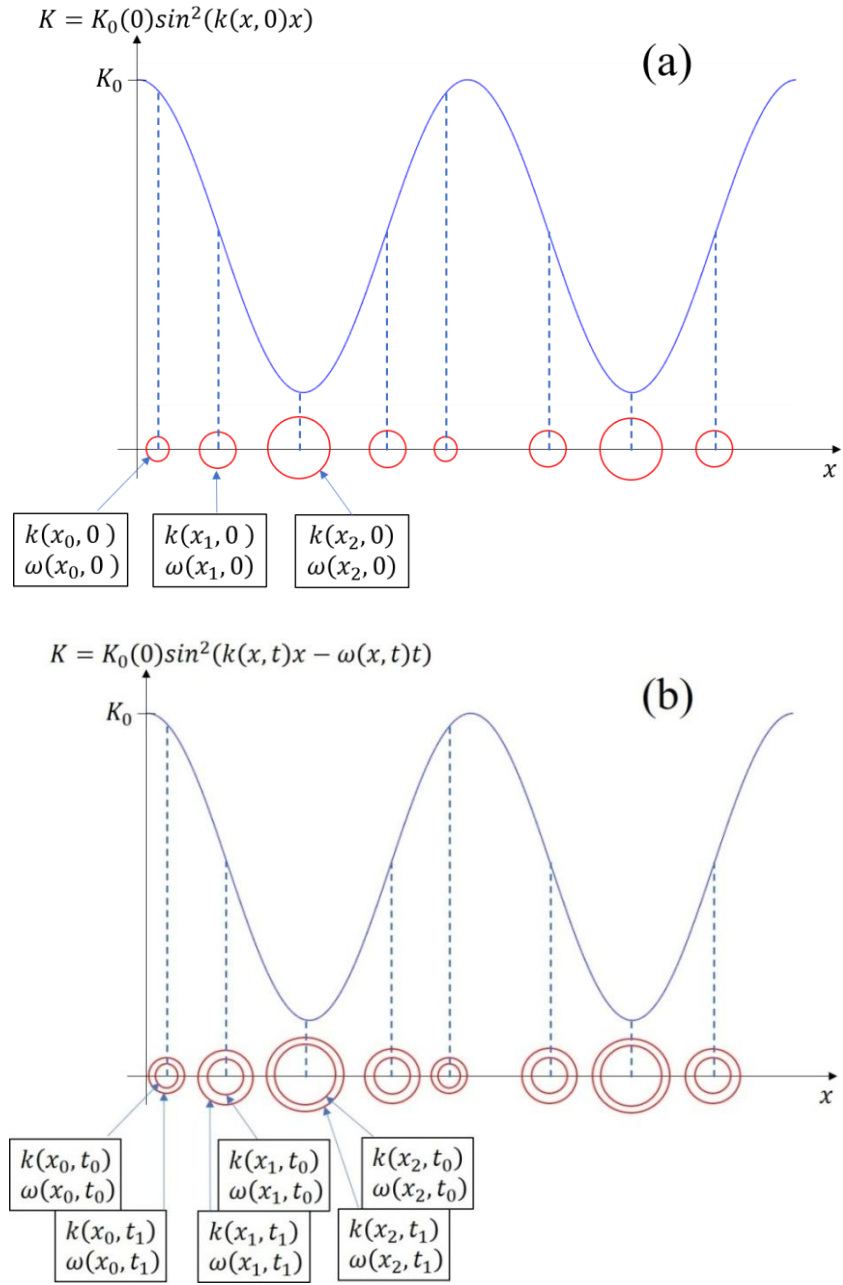

**FIG. 3.** Graphical representation of the curvature function: (a) $K = K_0(0)\sin^2\left(k(x,0)x\right)$ when $t=0$, and **(b)** $K = K_0(0)\sin^2\left(k(x,t)x-\omega(x,t)t\right)$ when $t\neq 0$. The form of the sinusoidal curve is only illustrative due that $k$ and $\omega$ depend on $x$ and $t$ and it is impossible to know the form of the real curve.

## 4. Simple unification of the Newton and Coulomb Laws

Consider the gaussian curvature of the sphere given by Eq. (1) which in turn satisfies Eq. (3a) with solution given by Eq. (3b). Then the gaussian curvature of the sphere can be written as

$$K_S = K_S^0(a)\sin^2(k_S x - \omega_S t) \tag{4}$$

From Eqs. (1) and (4), its amplitude is given by

$$K_S^0(a) = \frac{1}{r^2 \sin^2(k_S x - \omega_S t)} \tag{5}$$

where $k_S$ and $\omega_S$ are the wave number and frequency of the gaussian curvature function of the sphere.

In a similar way, as it was for the sphere, it is for the torus with negative gaussian curvature $K_T$. Then

$$K_T = K_T^0(a)\sin^2(k_T x - \omega_T t) \tag{6}$$

where $k_T$ and $\omega_T$ are the wave number and frequency of the gaussian curvature function of the torus.

From Eqs. (2) and (6), we also obtain its amplitude given by,

$$K_T^0(a) = \frac{\cos u}{r(a + r\cos u)\sin^2(k_T x - \omega_T t)} \tag{7}$$

From Eqs. (5) and (7) we have that

$$\frac{K_S^0(a)}{K_T^0(a)} = \frac{r(a + r\cos u)\sin^2(k_T x - \omega_T t)}{r^2 \cos u \sin^2(k_S x - \omega_S t)} \tag{8}$$

If a topological change and an annihilation process is going to happen, then we must have $a \to 0$.

This means that

$$\lim_{a \to 0} \frac{K_S^0(a)}{K_T^0(a)} = 1 \tag{9}$$

from the right side of Eq. (8) we have

$$\lim_{a \to 0} \frac{r(a + r\cos u)\sin^2(k_T x - \omega_T t)}{r^2 \cos u \sin^2(k_S x - \omega_S t)} = \frac{\sin^2(k_T x - \omega_T t)}{\sin^2(k_S x - \omega_S t)} \quad (10)$$

By taking limit both sides of Eq. (8) as $a$ approaches zero and taking into account Eqs. (9) and (10) we have

$$\frac{\sin^2(k_T x - \omega_T t)}{\sin^2(k_S x - \omega_S t)} = 1 \quad (11)$$

Since equation (11) means that an annihilation process took place, then we must have

$$k_S = k_T \quad , \quad \omega_S = \omega_T \quad (12a)$$

This means that there is an attraction between the two different membranes. But this is only a qualitative description and we need an operational formula as the inverse square law to display Newton and Coulomb formulas as a single one. This is only possible if in Eq. (11) we demand that, $k_S$, $k_T$, $\omega_S$ and $\omega_T$ depend on $x$ and $t$ instead of being them constant. Then, if Eq. (3b) satisfies Eq. (3a), the following relations must be satisfied,

$$\left(\frac{\partial k}{\partial t} x - \frac{\partial^2 \omega}{\partial t^2} - \omega\right)^2 - v^2 \left(k + x\frac{\partial k}{\partial x} - t\frac{\partial \omega}{\partial x}\right)^2 = 0$$

$$x\left(\frac{\partial^2 k}{\partial t^2} - v^2 \frac{\partial^2 k}{\partial x^2}\right) + t\left(v^2 \frac{\partial^2 \omega}{\partial x^2} - \frac{\partial^2 \omega}{\partial t^2}\right) - 2\left(v^2 \frac{\partial k}{\partial x} + \frac{\partial \omega}{\partial t}\right) = 0$$

for all $(x, t)$.

Equations (12a) tell us that Eq. (11) is valid for all $(x, t)$. In order of determining the way $(k_S - k_T) \to 0$ along the $x$-axis to fulfill Eq. (12a) when $\omega_S = \omega_T = \omega$, Eq. (11) implies that

$$\sin(k_T x - \omega t) = \sin(k_S x - \omega t + \sigma_1) \quad (12b)$$

with $\sigma_1 \neq 0$. Equation (12b) implies that,

$$k_T x - \omega t = k_S x - \omega t + \sigma_1 \quad (12c)$$

From which

$$k_S(x,t) - k_T(x,t) = \frac{\sigma_1}{x} \tag{12d}$$

We can also obtain this last relation, by deriving Eq. (11) with respect to $x$ and have

$$\cos(k_S x - \omega_S t)\sin(k_T x - \omega_T t)\left(x\frac{\partial k_S}{\partial x} + k_S - t\frac{\partial \omega_S}{\partial x}\right) - \sin(k_S x - \omega_S t)\cos(k_T x - \omega_T t)\left(x\frac{\partial k_T}{\partial x} + k_T - t\frac{\partial \omega_T}{\partial x}\right) = 0 \tag{13}$$

From which

$$\frac{\sin(k_T x - \omega_T t)}{\sin(k_S x - \omega_S t)} = \frac{\left(x\frac{\partial k_T}{\partial x} + k_T - t\frac{\partial \omega_T}{\partial x}\right)\cos(k_T x - \omega_T t)}{\left(x\frac{\partial k_S}{\partial x} + k_S - t\frac{\partial \omega_S}{\partial x}\right)\cos(k_S x - \omega_S t)} \tag{14}$$

From Eq. (11) we have

$$\frac{\sin(k_T x - \omega_T t)}{\sin(k_S x - \omega_S t)} = \pm 1 \tag{15}$$

By replacing Eq. (15) in (14) in the case when the right member from Eq. (15) is +1, we have that

$$1 = \frac{\left(x\frac{\partial k_T}{\partial x} + k_T - t\frac{\partial \omega_T}{\partial x}\right)\cos(k_T x - \omega_T t)}{\left(x\frac{\partial k_S}{\partial x} + k_S - t\frac{\partial \omega_S}{\partial x}\right)\cos(k_S x - \omega_S t)} \tag{16}$$

We simplify Eq. (16) to obtain

$$1 = \frac{x\frac{\partial k_T}{\partial x} + k_T - t\frac{\partial \omega_T}{\partial x}}{x\frac{\partial k_S}{\partial x} + k_S - t\frac{\partial \omega_S}{\partial x}} \tag{17a}$$

with

$$\frac{\cos(k_T x - \omega t)}{\cos(k_S x - \omega t)} = 1 \tag{17b}$$

But (17b) imply that

$$\cos(k_T x - \omega t) = \cos(k_S x - \omega t + \sigma_2) \tag{17c}$$

and from Eq. (17c) we have again that

$$k_T x - \omega t = k_S x - \omega t + \sigma_2 \tag{17d}$$

From which

$$k_S(x,t) - k_T(x,t) = \frac{\sigma_2}{x} \tag{17e}$$

Now, from Eq. (17a)

$$x\left(\frac{\partial k_S}{\partial x} - \frac{\partial k_T}{\partial x}\right) - t\left(\frac{\partial \omega_S}{\partial x} - \frac{\partial \omega_T}{\partial x}\right) = k_T - k_S \tag{18}$$

In a similar way by deriving Eq. (11) with respect to $t$, we have

$$x\left(\frac{\partial k_S}{\partial t} - \frac{\partial k_T}{\partial t}\right) - t\left(\frac{\partial \omega_S}{\partial t} - \frac{\partial \omega_T}{\partial t}\right) = \omega_S - \omega_T \tag{19}$$

We can also obtain Eq. (12d) or what is the same Eq. (17e) by solving Eqs. (18) and (19) with $\omega_S = \omega_T$. In fact, from Eq. (18) we have

$$x\left(\frac{\partial k_S}{\partial x} - \frac{\partial k_T}{\partial x}\right) = k_T - k_S \tag{20}$$

and from Eq. (19)

$$\frac{\partial k_S}{\partial t} - \frac{\partial k_T}{\partial t} = 0 \tag{21}$$

By integrating Eq. (21), we have that

$$k_S - k_T = f(x) \tag{22}$$

By deriving Eq. (22) with respect to $x$, we get

$$\frac{\partial k_S}{\partial x} - \frac{\partial k_T}{\partial x} = f'(x) \tag{23}$$

By replacing Eqs. (23) and (22) in Eq. (20)

$$xf'(x) + f(x) = 0 \tag{24}$$

By solving this last differential equation, we have

$$f(x) = \frac{\sigma}{x} \tag{25}$$

Taking into account Eq. (22) in (25), we have

$$k_T(x,t) - k_S(x,t) = -\frac{\sigma}{x} \tag{26}$$

Which is the same equation (12d) or (17e) with $\sigma_1 = \sigma_2 = \sigma$.

Since that in Eq. (3b) $k = \dfrac{\pi}{\lambda}$ becomes a wave number with fundamental spatial period equal to $\dfrac{\pi}{k}, \dfrac{\xi\pi}{k}$ is also a period. Then $k = \dfrac{\xi\pi}{\lambda}$ is also a wavenumber for the same Eq. (6), and $\xi$ is a dimensionless parameter to be determined. Therefore, from Eqs. (4) and (6), we have

$$k_S = \frac{\xi_S \pi}{\lambda_S} \text{ y } k_T = \frac{\xi_T \pi}{\lambda_T} \tag{27}$$

From which

$$k_S - k_T = \frac{\xi_S \pi}{\lambda_S} - \frac{\xi_T \pi}{\lambda_T} \tag{28}$$

By using the de Broglie relation $\lambda = \dfrac{h}{p}$ in Eq. (28), we have that

$$k_S - k_T = \frac{\xi_S \pi}{\dfrac{h}{p_S}} - \frac{\xi_T \pi}{\dfrac{h}{p_T}} \tag{29}$$

From which

$$k_S - k_T = \frac{\pi}{h}\left(\xi_S p_S - \xi_T p_T\right) = \frac{\beta}{x} \tag{30}$$

If $\xi_S = \xi_T = \xi$ in order to obtain the wished expression for the potential such that by applying the gradient we get a conservative force, we obtain

$$\frac{\xi\pi}{h}\left(p_S - p_T\right) = \frac{\sigma}{x} \tag{31}$$

From Eq. (31), we have finally

$$p_S - p_T = \frac{h\sigma}{\xi\pi x} \tag{32a}$$

By multiplying Eq. (32a) by the light velocity $c$

$$p_S c - p_T c = \frac{hc\sigma}{\xi\pi x} \tag{32b}$$

Then,

$$V = E_S - E_T = \frac{hc\sigma}{\xi\pi x} \tag{33}$$

Where $E_S = p_S c$ and $E_T = p_T c$ are the corresponding energies associated to the sphere and Torus.

By deriving Eq. (33)

$$\frac{\partial V}{\partial x} = -\frac{hc\sigma}{\xi \pi x^2} \tag{34}$$

Which must have dimensions of force and then $\sigma$ must be dimensionless. Then, defining $F_{ST}$ as the force between the sphere and the torus, by

$$F_{ST} = -\frac{hc\sigma}{\xi \pi x^2} \tag{35}$$

where $\xi$ is an integer number because $\frac{\xi \pi}{k}$ is the spatial period, and then the force given by (35) is itself quantized. It will be shown that according to the value of $\xi$, Eq. (35) will describe the nuclear force[1], the electrostatic interaction force and the gravitational interaction force.

In order to calculate the value of $\xi$ for the Coulomb interaction $(\xi_C)$, we put $|F_{ST}| = F_C$, where $F_C$ is the Coulomb force between two electrons, i.e.,

$$\frac{hc\sigma}{\xi_C \pi x^2} = \frac{K_e e^2}{x^2} \tag{36}$$

---

[1] This formula result to be valid when is contrasted with the Yukawa potential given by $V(x) = \beta \frac{e^{-\mu x}}{x}$, where $\beta = \frac{hc}{2\pi} \alpha$, $\alpha$ the fine structure constant and $\mu = \frac{m_\pi c}{\hbar}$, $m_\pi$ is the mass of pion. For this we make $\frac{\frac{H}{x} - \beta \frac{e^{-\mu x}}{x}}{\beta \frac{e^{-\mu x}}{x}} = \varepsilon$ where $\frac{H}{x}$ is the potential of the Coulomb type and it is demanded that $\varepsilon \approx 0$. Then $H = e^{-\mu x}(\beta + \varepsilon \beta) \approx (1 - \mu x)(\beta + \varepsilon \beta)$, by Taylor expansion. Hence $H = A - Bx$ where $A = \beta(1+\varepsilon)$ and $B = \mu\beta(1+\varepsilon)$. Since we are interested $B \approx 0$ in order that $H$ be constant such that $\frac{H}{x} \approx \beta \frac{e^{-\mu x}}{x}$ we must have that $\varepsilon \approx 0$ and $0 < \mu\beta \ll 1$, and this is true.

On the left – hand side of Eq. (36) is assumed that the value of the elementary charge is 1 and, on the right – hand side $K_e = \dfrac{1}{4\pi \in_0}$. From Eq. (36) we have that,

$$\frac{2\sigma}{\xi_C} = \frac{e^2}{4\pi \in_0 \hbar c} \tag{37a}$$

But

$$\frac{e^2}{4\pi \in_0 \hbar c} = \alpha \tag{37b}$$

Then

$$\frac{2\sigma}{\xi_C} = \alpha \tag{37c}$$

$$\frac{\sigma}{\xi_C} = \frac{\alpha}{2} = 3.6486762846 \times 10^{-3} = \frac{0.9997373062}{274} \tag{37d}$$

From Eq. (37d) we have that $\sigma = 9997373062$ and $\xi_C = 274$. For simplicity we make $\sigma = 1$ and Eq. (35) can be written as

$$F_{ST} = -\frac{hc}{274\pi x^2} \tag{37e}$$

The corresponding uncertainty to $\xi_C$ and $\sigma$ separately is not possible to determine [2].

Hence, the force that corresponds to the interaction between two elementary electric charges (electrons), that will be designed by $F_{ST-C}$, is given by the equation,

$$F_{ST-C} = -\frac{hc}{274\pi x^2} \tag{38}$$

---

[2] Here, we can only put down the uncertainty in the value of $\dfrac{\sigma}{\xi_C}$. This is because $\dfrac{\sigma}{\xi_C} = \dfrac{\alpha}{2}$. Then $d\left(\dfrac{\sigma}{\xi_C}\right) = \dfrac{d\alpha}{2} = 0.0000000006 \times 10^{-3}$ which is less than that of $d\alpha = 0.0000000011 \times 10^{-3}$. This means that the value of $\dfrac{\sigma}{\xi_C}$ is a good representative of the true value of $\dfrac{\sigma}{\xi_C}$. For the calculations we used $\alpha = 7.2973525693 \times 10^{-3} \pm 0.0000000011 \times 10^{-3}$ CODATA (2018): physics.nist.gov/constants.

In this last equation no electrical charge is present [3].

But due to the charge quantization, the previous formula can be rewritten for the case of interaction between any two charged bodies $\Omega_1, \Omega_2$ by writing,

$$F_{ST-C} = (-1)^l \frac{hcn^{\Omega_1}n^{\Omega_2}}{274\pi x^2} \quad (39)$$

Where,

$n^{\Omega_1}$ : number of elementary charges in body $\Omega_1$

$n^{\Omega_2}$ : number of elementary charges in body $\Omega_2$

$l = \begin{cases} \text{even number if electric charges have same sign} \\ \text{odd number if electric charges have different sign} \end{cases}$

Eq. (39) is equivalent to the Coulomb's law formula.

It is observed that Eq. (39) is expressed in terms of the fundamental constants, such as the Planck's constant and the light velocity, besides being characterized by the value of the dimensionless parameter $\xi_C = 274$. In order to achieve the simple unification of the Newton's law and Coulomb's law of the gravitational and electrostatics interaction, we must obtain the value of an elementary mass.

For this, it is first assumed that the nuclear force designated by $F_{ST-N}$, is described by Eq. (35) for some value of $\xi_N$. To determine the value of $\xi_N$ we consider the relative intensity between the nuclear force and the electrostatic interaction force

$$\frac{F_{ST-N}}{F_C} = \frac{\frac{h\sigma c}{\xi_N \pi x^2}}{\frac{e^2}{4\pi \epsilon_0 x^2}} = \frac{2\sigma}{\xi_N \alpha} \quad (40a)$$

---

[3] Eq. (38) can be obtained from the fine structure constant written as $\alpha = \frac{e^2}{4\pi \epsilon_0 \hbar c}$ in MKS units. The form here deduced let us interpret $\frac{\alpha}{2}$ as $\frac{1}{\xi}$ where $\xi = 274$ is such that $\frac{\xi \pi}{k}$ is the period of the solution (3b) of equation (3a).

Then, we use the relative intensity between the nuclear interaction force and the electrostatic interaction force as equal to the value of the fine structure constant $\alpha^{-1}$, the electromagnetic coupling constant defined from the Coulomb law for heavy nonrelativistic particles.[12] Obviously that he value of $\alpha^{-1}$ differs from that used at a scale of the order of Z boson mass $M_Z$ equal to $128.93 \pm 0.06$.[13] Then,

$$\frac{2\sigma}{\xi_N \alpha} = \alpha^{-1} \tag{40b}$$

$$\frac{\sigma}{\xi_N} = \frac{1}{2} \tag{40c}$$

Then $\xi_N = 2$ and again $\sigma = 1$

Hence,

$$F_{ST-N} = \frac{hc}{2\pi x^2} \tag{41}$$

Which is the formula for the nuclear interaction force for at least two entities[4].

At the same way, the relative intensity between the nuclear interaction force and the Newton's gravitational interaction force $\left(F_G = \frac{GM^2}{x^2}\right)$ is $5.7099 \times 10^{44}$, a good value estimated by the authors for the purpose of the present paper [5], which allow us to determine the value of the elementary mass $M$. Hence

$$\frac{F_{ST-N}}{F_G} = \frac{\frac{hc}{2\pi x^2}}{\frac{GM^2}{x^2}} = 5.7099 \times 10^{44} \tag{42}$$

from which

---

[4] For $\xi_N = 2$, Eq. (33) becomes $\frac{hc}{2\pi x}$. When $0 < x \leq 1.5 \times 10^{-15} m$, we have $\frac{2}{6\pi} \times 10^{15} hc \leq \frac{hc}{2\pi x}$ or $131.823 \leq \frac{19.773 \times 10^{-14}}{x}$ in units of MeV. This last inequality involves the masses of all hadrons in the range of $0 < x \leq 1.5 \times 10^{-15} m$. This fact lead us to trust, that Eq. (41) can be used as a formula for the nuclear force seen as a central force. The entities that interact here remain to be explored.

[5] This value has been adjusted in such a way that the value of the elementary mass corresponds to the mass of the electron. In current literature, this value is of the order of $10^{38}$ or $10^{39}$. For two quarks separated $3 \times 10^{-17} m$ the intensity is of the order of $10^{43}$ (http://uw.physics.wisc.edu). To greater fundamental particles and shorter separations, the greater the intensity. Hence, $5.7099 \times 10^{44}$ is an adequate value.

$$M = 9.108174007326300 \times 10^{-31} kg \qquad (43)$$

which is very closely to the value of the mass of the electron. From now on we will consider the mass of the electron as an elementary mass. This last value let us calculate the value of $\xi_G$ for the gravitational interaction force, by analogy to the electrostatic case, we put $F_{ST-G} = F_G$, such as

$$\frac{hc\sigma}{\xi_G \pi x^2} = \frac{GM^2}{x^2} \qquad (44)$$

from which

$$\frac{\sigma}{\xi_G} = 8.75904697010764 \times 10^{-46} = \frac{1}{1.14198 \times 10^{45}} \qquad (45)$$

In a similar way, as previously, the corresponding uncertainty to $\xi_G$ and $\sigma$ separately is not possible to determine [6].

Then $\xi_G = 1.14198 \times 10^{45}$ and again $\sigma = 1$.

Hence, our corresponding formula for the gravitational interaction between two elementary masses is

$$F_{ST-G} = -\frac{hc}{1.14198 \times 10^{45} \pi x^2} \qquad (46a)$$

---

[6] Here, we can only put down the uncertainty in the value of $\left(\frac{\sigma}{\xi_G}\right)$ which is calculated from $\left(\frac{\sigma}{\xi_G}\right) = \frac{GM^2 \pi}{hc}$ by taking its differential $\left(\frac{\sigma}{\xi_G}\right) = \left|\frac{M^2 \pi}{hc}\right| dG + \left|\frac{2GM\pi}{hc}\right| dM$. The value is $d\left(\frac{\sigma}{\xi_G}\right) = 0.00019680626453 \times 10^{-46}$. This means that the calculated value of $\left(\frac{\sigma}{\xi_G}\right)$ is a good representative of the true value of $\left(\frac{\sigma}{\xi_G}\right)$. For the calculations we have used $G = 6.67430 \times 10^{-11} \pm 0.00015 \times 10^{-11} m^3 kg^{-1} s^{-2}$ and $M = m_e = 9.1093837015 \times 10^{-31} \pm 0.0000000028 \times 10^{-31} kg$. CODATA: physics.nist.gov/constants.

where the two elementary masses are absorbed and no mass is present. This means that Eq. (44) can be written as,

$$\frac{hc}{1.14198 \times 10^{45} \pi x^2} = G \frac{m_e^2}{x^2} \tag{46b}$$

Where the absorbed elementary masses agree with the value given by Eq. (43). From now on for the calculations we will consider $M = m_e$ since the percentage error between $M$ and $m_e$ is around 0.0133%.

For any two masses $\Omega_1$ and $\Omega_2$ and in order to keep the similarity with Eq. (39), we add to Eq. (46a) $n^{\Omega_1} n^{\Omega_2} (-1)^l$ and we write it as

$$F_{ST-G} = (-1)^l \frac{hc n^{\Omega_1} n^{\Omega_2}}{1.14198 \times 10^{45} \pi x^2} \tag{47}$$

Where

$n^{\Omega_1}$ : number of elementary masses in body $\Omega_1$

$n^{\Omega_2}$ : number of elementary masses in body $\Omega_2$

$l = \begin{cases} \text{even number if masses have different sign} \\ \text{odd number if masses have same sign} \end{cases}$

Here, we write the values for $l$ in a different way in that was written for Eq. (39), based partially on the ordinary experience and that proposed by Farnes.[14] He proposed gravitational attractions for masses of equal sign, and gravitational repulsion for masses of different sign. However, this remains as an open problem in the context of this work. Eq. (47) is equivalent to the formula for the Newton's gravitational interaction. In fact, let us see that this is right.

Every material body is made up of atoms, and these of electrons, protons and neutrons. According to nuclear physics the mass of a nucleus is less than the sum of the masses of its components, because the missing mass (mass defect) has become in the binding energy of the nucleons. But general theory of relativity tell us that any form of energy is a source of gravity as the rest mass is. Then, we do not take the mass defect into account and the masses $m_1$ and $m_2$ of the two any bodies can be expressed as

$$m_i = p_i m_p + n_i m_n + e_i m_e; \quad i = 1, 2 \tag{48a}$$

Where

$p_i$ : number of protons

$n_i$ : number of neutrons

$e_i$ : number of electrons

If :

$n_p^e$ : exact number of electrons in a proton

$n_n^e$ : exact number of electrons in a neutron

$\varepsilon_{p<e}$ : proper fractional number corresponding to a part of a proton such that $0 < \varepsilon_{p<e} m_p < m_e$

$\varepsilon_{n<e}$ : proper fractional number corresponding to a part of a neutron such that $0 < \varepsilon_{n<e} m_n < m_e$

We can write

$$m_i = p_i \left( \left( n_p^e + \varepsilon_{p<e} \right) m_e \right) + n_i \left( \left( n_n^e + \varepsilon_{n<e} \right) m_e \right) + e_i m_e \; ; 0 < \varepsilon_{p<e} m_p < m_e \; ; 0 < \varepsilon_{n<e} m_n < m_e \tag{48b}$$

$$m_i = \left( p_i n_p^e + n_i n_n^e + e_i \right) m_e + \left( p_i \varepsilon_{p<e} + n_i \varepsilon_{n<e} \right) m_e \; ; 0 < \varepsilon_{p<e} m_p < m_e \; ; 0 < \varepsilon_{n<e} m_n < m_e \tag{48c}$$

Hence

$$m_1 m_2 = \left( \left( p_1 n_p^e + n_1 n_n^e + e_1 \right) \left( p_2 n_p^e + n_2 n_n^e + e_2 \right) \right) m_e^2 + R\left( \varepsilon_{p<e}, \varepsilon_{n<e} \right) m_e^2 \tag{48d}$$

With

$$R\left( \varepsilon_{p<e}, \varepsilon_{n<e} \right) = \left( \left( p_1 \varepsilon_{p<e} + n_1 \varepsilon_{n<e} \right) \left( p_2 n_p^e + n_2 n_n^e + e_2 \right) + \left( p_2 \varepsilon_{p<e} + n_2 \varepsilon_{n<e} \right) \left( p_1 n_p^e + n_1 n_n^e + e_1 \right) \right. \tag{48e}$$
$$\left. + \left( p_1 \varepsilon_{p<e} + n_1 \varepsilon_{n<e} \right) \left( p_2 \varepsilon_{p<e} + n_2 \varepsilon_{n<e} \right) \right) \geq 0$$

From Eq. (48d) we deduce

$$\left( \left( p_1 n_p^e + n_1 n_n^e + e_1 \right) \left( p_2 n_p^e + n_2 n_n^e + e_2 \right) \right) m_e^2 \leq m_1 m_2 \tag{48f}$$

Since

$$0 < R\left( \varepsilon_{p<e}, \varepsilon_{n<e} \right) \square \left( \left( p_1 n_p^e + n_1 n_n^e + e_1 \right) \left( p_2 n_p^e + n_2 n_n^e + e_2 \right) \right) \tag{48g}$$

Eq. (48f) must be really rewritten as

$$\left(\left(p_1 n_p^e + n_1 n_n^e + e_1\right)\left(p_2 n_p^e + n_2 n_n^e + e_2\right)\right) m_e^2 \approx m_1 m_2 \tag{48h}$$

By using Eq. (46b) and taking into account Eq. (48h), we can write

$$\frac{hc\left(p_1 n_p^e + n_1 n_n^e + e_1\right)\left(p_2 n_p^e + n_2 n_n^e + e_2\right)}{1.14198 \times 10^{45} \pi x^2} = G\frac{\left(p_1 n_p^e + n_1 n_n^e + e_1\right)\left(p_2 n_p^e + n_2 n_n^e + e_2\right) m_e^2}{x^2} \approx G\frac{m_1 m_2}{x^2} \tag{48i}$$

$$\frac{hc n^{\Omega_1} n^{\Omega_2}}{1.14198 \times 10^{45} \pi x^2} \approx G\frac{m_1 m_2}{x^2} \tag{48j}$$

Where $n^{\Omega_i} = n_p^e p_i + n_n^e n_i + e_i$; $i = 1, 2$; expresses the number of electrons in the body of mass $m_i$ and then the number of elementary masses $n^{\Omega_i}$. Eq. (48f) tell us really that the left-hand side of Eq. (48j) approximates the right-hand side by defect.

From the last two numerators of Eq. (48i) we deduce that

$$m_i \approx n^{\Omega_i} m_e \; ; i = 1, 2 \; ; \tag{49}$$

Table 1 shows how the mass defect intervenes in the calculation of the gravitational force for the case of very small masses such as nucleus of some isotopes in using both the formula $F_G$ and the formula $F_{ST-G}$ deduced here, taking for the calculations $n_p^e = 1836$ and $n_n^e = 1838$. To show this, we have considered the nucleus of the isotopes: Beryllium – 9 versus Sodium - 23 and Aluminium – 27 versus Cobalt – 59. It is found that the resultant value for $F_{ST-G}$ has a percentage error $\varepsilon[\%]$ of the order of $1 \times 10^{-2} \%$ with respect to $F_G$ calculated when the mass defect (MD) was leaving out in the calculation of $F_G$ (Without – Mass Defect: WO-MD). Say, when the masses for calculating $F_G$ are given by Eq. (48a). On the other hand, when the mass defect was considered (With – Mass Defect: W-MD) in the calculation of $F_G$, the percentage error $\varepsilon[\%]$ of $F_{ST-G}$ with respect to $F_G$ is of the order of magnitude zero. Say, when the masses for calculating $F_G$ are given by experimental data. Similarly, results were found in using the traditional formula of Coulomb's law $(F_C)$ with the $F_{ST-G}$ formula deduced here (tables are not shown).

**Table 1** Comparison of the Newton force, between two different isotopes, using the traditional formula $(F_G)$ and the deducted formula $(F_{ST-G})$. It is shown the percentage errors $\varepsilon[\%]$ for the calculations of $F_{ST-G}$ with respect to $F_G$ without (WO-MD) and with (W-MD) mass defect, where $\varepsilon[\%] = \left|\dfrac{F_{ST-G} - F_G}{F_G}\right| \times 100$. Data for the value of isotopes were found in Ref. 15 and the value of atomic mass unit $(1\,u = 1.6605387 \times 10^{-27}\,kg)$ on Ref. 16:

| Masses | | Distance (m) | $F_G(N)$ | | $F_{ST-G}(N)$ | $\varepsilon[\%]$ | $\varepsilon[\%]$ |
|---|---|---|---|---|---|---|---|
| | | | WO-MD | W-MD | | WO-MD | W-MD |
| Berillium-9 $p(4)$ $n(5)$ | Sodium-23 $p(11)$ $n(12)$ | $10^{-11}$ | 3.87094e-40 | 3.81301e-40 | 3.86806e-40 | 0.0742872 | 1.44373 |
| Aluminium-27 $p(13)$ $n(14)$ | Cobalt-59 $p(27)$ $n(32)$ | $10^{-13}$ | 2.97887e-35 | 2.92638e-35 | 2.97667e-35 | 0.0738139 | 1.71871 |

It is observed that formula (47) is the same formula (39) except that the value of the dimensionless parameter $\xi_G$ is equal to $1.14198 \times 10^{45}$.

## 5. Results and Discussions

Table 1 shows that the deduced formula (47) for $F_{ST-G}$, taking into account the hypothesis based on membranes, provides similar results in the calculations of gravitational force as it is given by the traditional formula of Newton's law $F_G$, when the mass defect is not considered (WO-DM) and when the mass defect is considered (W-MD) in the calculation of $F_G$. In both cases, formulas $F_{ST-G}$ and $F_G$ are in good agreement. Similar results were found comparing results obtained by Eq. (39) for $F_{ST-C}$ and that found with the traditional formula of Coulomb's law $(F_C)$ (tables are not shown here by the reason given on the footer 3). Hence, formulas given by Eqs. (39) and (47) are validated, and due to fact that both formulas have been deduced from Eq. (35), these formulas can be written in a unique general formula given by

$$F = (-1)^l \frac{hcn^{\Omega_1} n^{\Omega_2}}{\xi \pi x^2} \qquad (50)$$

such that if the dimensionless parameter $\xi = 274$, $F$ corresponds to the Coulomb's law of electrostatic interaction $F_{ST-C}$, and if $\xi = 1.14198 \times 10^{45}$, $F$ corresponds to the Newton's law of universal gravitation $F_{ST-G}$ as shown in Fig. 4. The value of $\xi$ is such that $\frac{\xi \pi}{k}$ is the spatial period of the solution (Eq. (3b)) of the wave equation (Eq. (3a)). Furthermore, the classical proportionality constants for the Coulomb's law $K_e$ and Newton's law $G$ are replaced by the constants $\frac{hc}{274\pi}$ and $\frac{hc}{1.14198 \times 10^{45} \pi}$ respectively, which have the same units, and no more product of electric charges and product of masses, in these new formulas, as it was the case in the classical formulas that made the difference. Now, both forces are of the same nature in the sense that both are the result of the interaction of the membranes postulated above. On the other hand, an important feature from this formula is that, it is written in terms of the speed of light in the vacuum, which is often associated with relativistic phenomena, and the Planck's constant, which is normally associated with quantum phenomena.

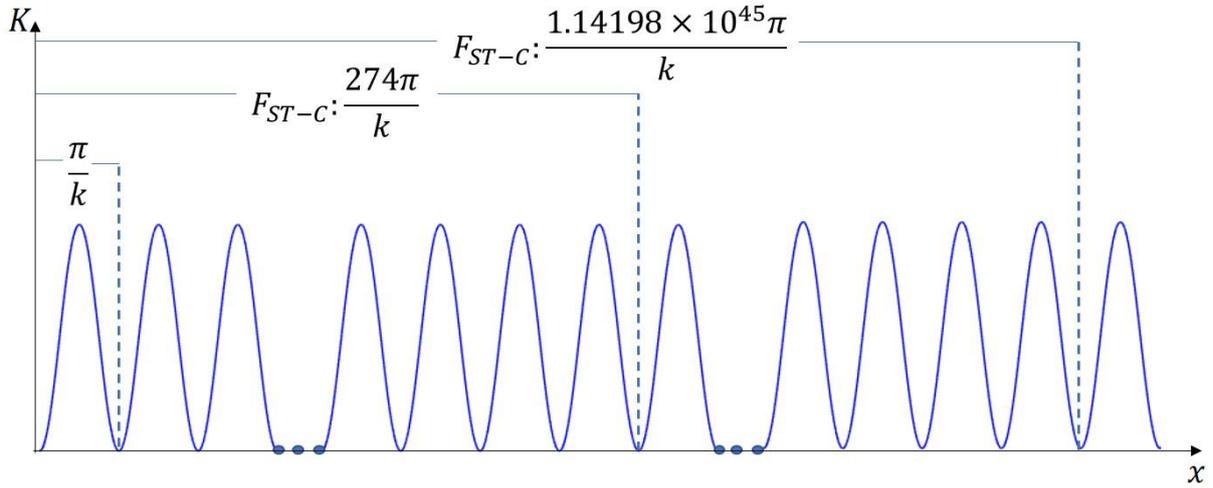

**FIG. 4.** Periods of the curvature function $K = K_0 \sin^2(kx - \omega t)$ which make the difference between the forces $F_{ST-C}$ and $F_{ST-G}$. $\frac{\pi}{k}$ is the fundamental period and $k$ the wave number. The form of the sinusoidal curve is only illustrative.

According to Eqs. (33) and (34) the above forces are the result of the rate of change with the distance of the difference of energies $E_S$ and $E_T$ associated to oscillating membranes we have postulated, when the angular frequency $\omega_S$ equals the angular frequency $\omega_T$, which means that the particles should be in resonance to able to interact between them.

It is worth to emphasize that several authors have tried to unify both Coulomb's law and Newton's law without finding a satisfactory general formula. For instance, Caillon[17] derived a formula of possible unification by considering electric charge as the fourth component of the particle momentum in five dimensional space – time which was compactified on a circle with an extremely small radius determined from the fundamental physics constants. The formula obtained by Caillon is,

$$\vec{F} = -\frac{GM_1M_2}{r^2}\vec{u} + \frac{q_1q_2}{4\pi\epsilon_0 r^2}\vec{u}$$

This result is the sum of the classical versions of the formulas for the Coulomb's law and Newton's law and is not a formula with a unique term, as it is in our case. Furthermore, nothing tells us about the relation between the natures of the two forces expressed by the Coulomb's law and Newton's law. On the other hand, Gomez et al.[16] and Li,[17] in describing a possible unification theory of electromagnetism and gravity doesn't show an unification formula for the Coulomb's law and Newton's law. It is worth emphasizing that Eq. (35) leaves as an open problem to clarify the nature of the attractive and repulsive force of Coulomb's law, as well as the possible existence of positive and negative masses and then an attractive and repulsive force according to Newton's law, from the point of view of the postulated membranes.

As it is well known, the elementary charge corresponding to that of the electron measured in terms of the Planck charge is given by $e = \sqrt{\alpha}\, q_p$ where $\alpha$ and $q_p$ are the fine $-$ structure constant and Planck charge respectively, from which $\alpha = \left(\dfrac{e}{q_p}\right)^2 \approx \dfrac{1}{137.03599911}$. The elementary mass $m_e$ of the electron can be written as $m_e = \sqrt{\alpha_G}\, m_p$ where $\alpha_p$ and $m_p$ are the coupling constant for the gravitational interaction and $m_p$ the Planck mass, from which $\alpha_G = \left(\dfrac{m_e}{m_p}\right)^2 \approx 1.751751596 \times 10^{-45}$. Although the relationship between $\alpha_G$ and gravitation is somewhat analogous to that of the fine-structure constant $\alpha$ and electromagnetism, we can see that whereas $\alpha$ is a function of the elementary charge which is a quantum that is independent of the choice of the particle, $\alpha_G$ is a function of the electron rest mass, being in this last case an arbitrariness. According to above $\alpha$ and $\alpha_G$ come from different origins. However, in the present work, the fine $-$ structure constant and its analogous $\alpha_G$ can be obtained from a function of the dimensionless parameter $\xi$, $\alpha(\xi) = \dfrac{2}{\xi}$, for $\xi = 274$ and $\xi = 1.14198 \times 10^{45}$ respectively. Furthermore, $\alpha = 1$ when $\xi = 2$ corresponding to that we called nuclear force (Eq. 41). Emphasizing that $\xi$ is such that $\dfrac{\xi \pi}{k}$ is the spatial period of the oscillating membranes postulated above.

Since the left hand side of Eq. (48j) approximates the right hand side by defect, and if in Eq. (48a) we take into account the mass defect, we hope that Eq. (48j) approximate more to a equality.

With the aim of showing the validity of formula (33), we are going to show that this equation can also be derived from the Yukawa's free –space equation[18]

$$\left(\nabla^2 - \frac{\partial^2}{c^2 \partial t^2} - \frac{1}{a^2}\right) U(\vec{r}, t) = 0 \tag{51}$$

where $a$ is a range parameter $(\Box\, 2\, fm)$. Equation (51) is certainly relativistically invariant and corresponds to the non-static case. Although so far $U$ is still a classical field, Yukawa took the decisive step of treating $U$ quantum mechanically, by looking for a propagating wave solution of the above equation in the form

$$U = e^{\left(\frac{i\vec{p}\cdot\vec{r}}{\hbar} - \frac{iEt}{\hbar}\right)} \tag{52}$$

In the one – dimensional space case, Eq. (51) becomes

$$\left(\frac{\partial^2}{\partial x^2} - \frac{\partial^2}{c^2 \partial t^2} - \frac{1}{a^2}\right) U(x, t) = 0 \tag{53}$$

and (52)

$$U = e^{\left(2\pi i \left(\frac{x}{\lambda} - \frac{t}{T}\right)\right)} \tag{54}$$

By replacing $\xi$ instead of number 2 in Eq. (54), we have

$$U = e^{\left(\xi\pi i\left(\frac{x}{\lambda} - \frac{t}{T}\right)\right)} = e^{\left(\frac{\xi i}{2}\left(\frac{px}{\hbar} - \frac{Et}{\hbar}\right)\right)} \tag{55}$$

By replacing Eq. (55) in Eq. (53), one finds that

$$E = \sqrt{p^2 c^2 + \frac{4 c^2 \hbar^2}{\xi^2 a^2}} \tag{56}$$

Comparing Eq. (56) with the standard relation for a massive particle in special relativity given by

$$E = \sqrt{p^2 c^2 + m^2 c^4}$$

We find that

$$mc^2 = \frac{hc}{\xi \pi a} \quad (57)$$

By putting $E = mc^2 = V$ in Eq. (57) an replacing the range parameter $a$ by $x$, we finally have

$$V = \frac{hc}{\xi \pi x} \quad (58)$$

As long as $a$ and $x$ are the same thing. Furthermore, Eq. (57) tell us that

$$m = \frac{h}{\xi \pi a c}$$

Is the mass of the quantum of the finite – range force field $U$. Then, Eq. (57) is the same to Eq. (33).

## 6. Conclusions

In this work, a simple way of unifying the forces given in Coulomb's law and Newton's law of the universal gravitation was achieved taking into account an approach based on membranes. This simple unification was possible because both forces resulted to be of the same nature since their new corresponding formulas ($F_{ST-C}$ and $F_{ST-G}$) were derived from the interaction between the same membranes postulated in this work and under the same mechanism, being the link of unification the dimensionless parameter $\xi$. According to equations (33) and (34) the above forces are the result of the rate of change with the distance of the difference of energies $E_S$ and $E_T$ associated to oscillating membranes we have postulated and, when the angular frequency $\omega_S$ equals the angular frequency $\omega_T$. These formulas only differentiate on the dimensionless parameter $\xi$ present in both formulas, for which $\frac{\xi \pi}{k}$ is the spatial period of the solution of the wave equation that governs the oscillatory behavior of the membranes. This dimensionless parameter is like a dial of a radio. It let us tune the above forces. If $\xi = 274$ we have the Coulomb's law of electrostatic interaction, and if $\xi = 1.14198 \times 10^{45}$ we have the force for the Newton's law of universal gravitation. The constants $K_e$ and $G$ are no longer more present in the rewritten formulas of the forces. So, we have reformulated the four constants $h, c, G$ and $\alpha$ into four new constants $h, c, \xi_G, \xi_C$. However, in the first ones only $\alpha$ is dimensionless and in ours, two are dimensionless $(\xi_G, \xi_C)$. In this sense, we have reduced the number of constants with dimensions from three to only two. This fact let us unification, since by one hand, no more mass nor charge appear in the constants presents in the corresponding formulas for the forces, and on the other hand, the force in both cases are the same except that they have different values for $\xi$.

Even more, the possible unified formula is now quantized due that the value of $\xi$ is an integer number.

The one – dimensional variable $x$ of Eq. (50), is essentially the range parameter $a$ related to the quantum of the finite – range force field with mass $m$, by $a = \dfrac{h}{\xi \pi m c}$ in Yukawa's theory. When $m \to 0$ and then the virtual quanta of exchanging are photons, the Yukawa's potential becomes $V(x) = \dfrac{\beta}{x}$ which is essentially the Coulomb's potential and from which we can obtain the Coulomb's law. Equivalent to this and according to the present theory, if $\xi = 274$, we have the potential $V(x) = \dfrac{hc}{274 \pi x}$ in which the product of the two elementary charges is just included. What is new at this point is that if $\xi = 1.14198 \times 10^{45}$ we have the potential $V(x) = \dfrac{hc}{1.14198 \times 10^{45} \pi x}$ in which the product of the two elementary masses is just included and from which we have deduced the corresponding formula for the gravitational interaction and equivalent to that given by Newton. The simple way of unifying the formulas for the Coulomb's law and Newton's law of the universal gravitation rests in this argumentation too. Additionally,

Given that the forces between masses and charged bodies are of radial nature (central forces), the above – mentioned theory should not be difficult to extend to the three – dimensional case. It is worth emphasizing that although the method used to deduce the possible unified formula is not of a high mathematical level, as required by current theories, the calculations obtained by the unified formula are in good agreement with the calculations obtained by the traditional formulas of Coulomb and Newton. Therefore, these results leave open the possibility of formulating a possible theory based on membranes, which would allow to unify all the fundamental forces of nature.

**Acknowledgements**

The authors would like to thank Universidad Nacional del Santa – UNS and Federal University of Pampa – Unipampa for its partial financial support.

**References**

[1] Nagashima, Y.: *Elementary Particle Physics. Volume 1, Quantum Field Theory and Particles*, 1st ed. Wiley-VCH (2010)

[2] H. C. Corben, *Phys. Rev.* **69**, 225 (1946). DOI: https://doi.org/10.1103/PhysRev.69.225

[3] E. S. Fradkin and A. A. Tseytlin, Phys. Rep. **119,** 233 (1985). DOI: https://doi.org/10.1016/0370-1573(85)90138-3

[4] A. Davidson and D. A. Owen, Phys. Lett. B **166,** 123 (1986). DOI: https://doi.org/10.1016/0370-2693(86)91361-4

[5] J. E. Brandenburg, IEEE Trans. Plasma Sci. **20**, 944 (1992). DOI: 10.1109/27.199556

[6] A. K. T. Assis, in *Adv. Electromagn. Found.Theory Appl.* WORLD SCIENTIFIC 314(1995). DOI: https://doi.org/10.1142/9789812831323_0010

[7] P. Ghose, "Unification of Gravity and Electromagnetism II A Geometric Theory," e- print arXiv:1408.2403v3

[8] P. Ghose, "Unification of Gravity and Electromagnetism I: Mach's Principle and

Cosmology", e-print arXiv:1408.2507v4

[9]E. G. Haug, "Unification of Gravity and Electromagnetism GravityElectroMagnetism A Probability Interpretation of Gravity," *e-print* viXra:1604.0208

[10]S. H. Hendi and A. Sheykhi, "Entropic Corrections to Coulomb's Law," Int J Theor Phys (2012) 51:1125–1136, DOI 10.1007/s10773-011-0989-2

[11]M.P. Do Carmo, *Differential Geometry of Curves and Surfaces* 1st ed. Prentice-Hall (1976).

[12]S.S. M. Wong,Introductory Nuclear Physics Second Edition Wiley - VHC. Pag.3 (2004)

[13]J.G.Korner, A.A. Pivovarov and K. Schilcher, "On the running electromagnetic coupling constant at $M_z$," Eur. Phys. J. C 9, 551–556 (1999). DOI 10.1007/s100529900061

[14]J. S. Farnes, Astron. Astrophys. **620,** A92 (2018). DOI: https://doi.org/10.1051/0004-6361/201832898

[15]E. H. Norman, B. C. Tyler, K. B. John, V. T. Lauren, B. Jacqueline, R. L. John, G. M. Peter, O. Glenda, R. Etienne, H. T. Dorothy, W. Thomas, E. W. Michael and Y. Shigekazu, Pure and Appl. Chem. **90** (12), 1833 (2018). DOI: https://doi.org/10.1515/pac-2015-0703

[16]M. Joseph and G. Jean, Radioactivity Radionuclides Radiation. Springer, Berlin, Heidelberg (2005). DOI: https://doi.org/10.1007/3-540-26881-2_2

[17] J. C. Caillon, Phys. Lett. A **382**, 3307 (2018). DOI: https://doi.org/10.1016/j.physleta.2018.09.005

[18]A. Torres-Gomez, K. Krasnov and C. Scarinci, Phys. Rev. D **83**, 25023 (20111). DOI: https://doi.org/10.1103/PhysRevD.83.025023

[19]LX. Li, Front. Phys. **11**, 110402 (2016). DOI: https://doi.org/10.1007/s11467-016-0588-z

[20]J. J. R. Aitchison and A. J. G. Hey, Gauge Theories in Particle Physics Vol. I. Third Edition. Pag. 31 (2003).